\documentclass[fleqn,10pt]{wlscirep}
\usepackage[utf8]{inputenc}
\usepackage[T1]{fontenc}

\usepackage[english]{babel}

\usepackage{color,soul,xcolor,balance,colortbl, graphicx,tabularx,enumitem,longtable,booktabs}

\usepackage{amssymb}
\usepackage{amsmath}
\usepackage{tabularx}
\usepackage{booktabs}
\usepackage{subcaption}
\usepackage{xcolor}
\usepackage{enumitem}
\usepackage{arydshln}
\usepackage{hyperref}
\usepackage{tcolorbox}
\usepackage{float}
\usepackage{multirow}
\usepackage{authblk}
\usepackage{lscape}

\newcommand{\Tr}{\textsf{Human}}
\newcommand{\Ct}{\textsf{LLM}}

\definecolor{MediumBlue}{RGB}{0,0,225}
\newcommand{\edit}[1]{{{#1}}}

\title{\edit{Communication Styles and Reader Preferences of LLM- and Human-authored COVID-19 Information Explanations: A Case Study}}

\author[1]{Jiawei Zhou\textsuperscript{\dag}}
\author[1]{Kritika Venkatachalam\textsuperscript{\dag}}
\author[2]{Minje Choi}
\author[3]{Koustuv Saha}
\author[1]{Munmun De Choudhury}
\affil[1]{School of Interactive Computing, Georgia Institute of Technology, Atlanta, GA, USA}
\affil[2]{Amazon,  Seattle, WA, USA}
\affil[3]{Siebel School of Computing and Data Science, University of Illinois Urbana-Champaign, Urbana, IL, USA}

\affil[$\dag$]{Authors contributed equally to this research\newline}

\begin{abstract}
\textbf{Background:} With the wide adoption of large language models (LLMs) in \edit{assisting professionals to create and share information}, it is essential to examine their alignment with human communication styles and values \edit{in achieving communication effectiveness and acceptability}. We situate this study within the context of \edit{explaining} fact-checked health information, \edit{with a focus on communication styles because they influence whether explanations and corrections are understood, trusted, and acted upon.}
Recent studies have explored the potential of LLM-based fact-checking, but communication style differences between LLMs and human fact-checkers and associated reader perceptions remain under-explored. In this light, our study evaluates the communication styles of LLMs, focusing on how their explanations differ from those of humans in three core components of health communication: information linguistic features, sender persuasive strategies, and receiver value alignments.
        
\textbf{Methods:} We compiled a dataset of 1498 \edit{misinformation claims and corresponding} explanations from authoritative fact-checking organizations. Using this dataset, we employed chain-of-thought prompting with zero-shot and few-shot variations to generate LLM fact-checking responses to inaccurate health information. We drew from health communication theories and categorized communication styles along linguistic, persuasion, and value-based dimensions and measured how closely the LLM-generated responses aligned with professional explanations. Then, we examined human preferences with 99 participants who were unaware of LLM involvement and rated randomized fact-checking article pairs with switching orders. 
        
\textbf{Results:} Our findings reveal that LLM-generated articles showed significantly lower scores in persuasive strategies, certainty expressions, and alignment with social values and moral foundations. However, human evaluation demonstrated a strong preference for LLM content, with over 60\% responses favoring LLM articles for clarity, completeness, and persuasiveness. \edit{This reader preference of LLMs was driven by the structured presentation, clarity, and neutral tone, which may foster an impression of completeness and professionalism, even when the underlying content may lack nuance or rigor. }
        
\textbf{Conclusion:} Our results suggest that LLMs' structured approach to presenting information may be more effective at engaging readers despite scoring lower on traditional measures of quality in health communication. This indicates both the potential and the limitations of utilizing LLMs in health communication or fact-checking workflows. 
        
\end{abstract}

\begin{document}

\flushbottom
\maketitle
%
%
\thispagestyle{empty}



\section*{Background}

With the widespread advancement and proliferation of Large Language Models (LLMs) and generative AI, these technologies are increasingly being used for co-writing and writing assistance across a range of domains---including health communication tasks. Recent studies have demonstrated their potential to generate customized responses and provide support for complex information needs in health-related contexts~\cite{kuroiwa2023potential, BENICHOU2023101456}. \edit{For instance, work has explored LLMs' potential in both promoting pro-vaccination messages~\cite{karinshak2023working} and combating misinformation or myths~\cite {joshi2024ensuring, mittal2025onine}.}
\edit{While factual correctness remains the non-negotiable foundation of health communication, accuracy alone does not guarantee that information will be understood or accepted~\cite{schiavo2013health}.} 

\edit{Instead, in practice,} effective health communication \edit{is hard to achieve and} goes beyond \edit{conveying accurate and credible information. Prior research has highlighted that the effectiveness of health communication also depends on communication styles: how messages are presented and whether they resonate with audiences' values and contexts~\cite{schiavo2013health, rashkin-etal-2017-truth, borah2022correcting, kreuter2004role}. Specifically, the way messages are presented, such as readability~\cite{sakai2013role, wittink2018patient}, certainty~\cite{jin2025certainty}, and cognitive processing cues~\cite{kumar2025cognitive, chin2018multi, kaye2017use}, shapes the success of health communication as it influences the degree to which information is understood and trusted. Beyond linguistic styles, persuasive strategies such as evidence presentation or emphasis on impacts can increase perceived trust and action change~\cite{rains2007psychological, albarracin2003persuasive}, while alignment with shared moral and social values has been linked to greater message uptake and sharing tendency~\cite{okuhara2025effectiveness, borah2022correcting, xiao2023story, kreuter2004role, yang2024sharing}.}

One particularly critical health communication task is explaining misinformation. 
\edit{As with any type of health communication, if not more so, factual accuracy is necessary but not sufficient on its own for effective misinformation rebuttal, because resistance can stem from psychological, social, or value-based factors~\cite{swire2020public, ecker2022psychological}.}
Expert fact-checking, as one of the crucial traditional methods of misinformation rebuttal, has long-established standards for verifying information~\cite{bautista2021healthcare, swire2020public, guo-automated-fc}. Recent studies in automated fact-checking systems, specifically leveraging LLMs, have shown promise in streamlining this process~\cite{tang2024minicheck, rashkin-etal-2017-truth, hoes_altay_bermeo_2023, zhang2023llmbasedfactverificationnews}. 
Although LLMs have shown technical promise, evidence is still needed to examine communication styles and reader acceptability of LLM-generated content in assessing its potential to assist \edit{professionals} in communicating health misconceptions.

To address the above gap, this work examines whether LLM-generated explanations diverge from human explanations in both linguistic styles and value alignment and whether these differences influence reader preferences. Situating this work in fact-checking false claims during the COVID-19 pandemic, a health crisis time during which opinions were polarized~\cite{hart2020politicization} and people were susceptible to vulnerability and rumors~\cite{tasnim2020impact}, we seek to answer the following research questions (RQs):
\begin{itemize}
    \item[\textbf{RQ1.}] If and how do communication styles differ between LLMs and human fact-checking? 
    \item[\textbf{RQ2.}] Is there a difference in human preference between LLMs and human-generated fact-checking, and why?
\end{itemize}

To address the RQs, we collected a dataset of 1,498 \edit{misinformation claims from prior work~\cite{saenz2021covid, Shahi_Nandini_2020} and retrieved the corresponding human-written fact-checking articles listed in the original misinformation datasets}. Using chain-of-thought prompting~\cite{wei2022chain}, we constructed prompts based on journalism information assessment guidelines~\cite{zhou2023synthetic} and generated LLM fact-checking articles using a set of state-of-the-art models: GPT-4~\cite{achiam2023gpt}, Llama-2-70B~\cite{touvron2023llama}, and Mistral-7B~\cite{jiang2023mistral7b} with zero-shot and few-shot variations. To evaluate how closely LLM-generated responses aligned with human communication, we operationalized communication styles across three key components in health communication: information, receiver, and sender~\cite{schiavo2013health}. Specifically, we analyzed 1) information linguistic features of certainty, cognitive processing, and readability, 2) receiver value alignments in social values and moral foundations, and 3) sender persuasive strategies. We found that LLM-generated fact-checking articles showed distinct differences in communication style compared to human writing. LLMs employed fewer persuasive strategies with lower language impact and evidence use, expressed less certainty, and scored lower on moral foundations for authority and fairness. They also demonstrated lower alignment with social values like tradition and conformity. However, this neutrality can help minimize cultural bias. While LLMs display more cognitive expressions, their higher readability scores suggest the use of more complex language.

Then, to explore human preferences for communication styles between human and LLM content, we recruited 99 participants to rate and provide feedback on the clarity, completeness, and persuasiveness of randomly selected fact-checking article pairs. Participants were not informed that some articles were AI-generated, and the order of article pairs was randomized to minimize bias. Our results showed a notable preference for LLM-generated content despite the limitations identified in linguistic analysis. Approximately 62\% of participants preferred LLM articles across all three dimensions -- clarity, completeness, and persuasiveness, for their focused presentation, objective and neutral tone, professional appearance, and language accessibility.

Together, our study provides implications for the potential use of LLMs in fact-checking and health communication. While LLMs demonstrate strengths in structured presentation and linguistic styles, their limitations in reasoning support, evidence engagement, and moral alignment raise concerns about their reliability in high-stakes domains. The contrasts between language analysis and human preferences suggest that perceived professionalism and clarity may overshadow deeper considerations of nuanced reasoning and holistic coverage. Future work should explore strategies to enhance LLMs' alignment with human reasoning and values while maintaining their communicative advantages.

\section*{Methods}

We retrieved \edit{COVID-19 misinformation claims from existing datasets and related} fact-checking articles written by authoritative fact-checking platforms.
\edit{Based on the misinformation claims,} we generated LLM fact-checking articles using chain-of-thought prompting. We compared the communication styles of the two datasets by extracting linguistic features, value alignments, and persuasive strategies. With approval from our Institutional Review Board, we then recruited 99 participants to assess the human preference in reasoning styles. 

\subsection*{\edit{Misinformation Claims and Human Fact-checking Articles}}

We retrieved \edit{both COVID-19 misinformation claims and related} fact-checking articles sourced from FNIR~\cite{saenz2021covid} and FakeNews~\cite{Shahi_Nandini_2020} datasets that span from February 2020 to July 2020. 
Datasets were \edit{selected based on the criteria that 1) they were} peer-reviewed, \edit{2) the determination of information veracity is annotated manually rather than through algorithms, and 3)} contained source links to fact-checking articles by authoritative fact-checking platforms. 
\edit{After collecting all false claims, we then scraped and retrieved corresponding full fact-checking articles using the source links in FNIR and FakeNews. These articles were written by expert and authoritative fact-checking organizations such as Poynter~\footnote{https://www.poynter.org/} and FactCheck~\footnote{https://www.factcheck.org/}}
After removing non-English content, the final human-written fact-checking dataset had a total of 1,498 articles \edit{to misinformation claims} (referred to as \Tr{}) with an average length of 765 (range [47-6473]).

\subsection*{LLM Fact-checking Articles}
\edit{Based on the retrieved misinformation claims,} we used a set of state-of-the-art models (GPT-4~\cite{achiam2023gpt}, Llama-2-70B~\cite{touvron2023llama}, and Mistral-7B~\cite{jiang2023mistral7b}) to generate LLM fact-checking articles. We employed chain-of-thought prompting~\cite{wei2022chain}, a widely adopted method known to assist LLMs in performing complex reasoning, with zero-shot and few-shot variations. In the zero-shot prompt, we did not present any examples of human fact-checked articles, which allows us to focus solely on the LLMs' independent reasoning process. For the few-shot prompt, we supplemented with examples of human fact-checked articles for both true and false claims. \edit{LLM outputs were generated using default API parameters (temperature = 1.0, top\_p = 1.0). We did not set a lower max\_tokens limit, as the defaults for all models exceeded the length of generations observed in both preliminary testing and the final dataset.}


To generate fact-checking articles that follow the same general journalism guidelines as human articles, we constructed our prompt based on an information assessment guideline~\cite{kovach2011blur, zhou2023synthetic} built upon prior research in journalism communication~\cite{kovach2011blur} and information credibility indicators~\cite{zhang2018structured}. 
\edit{This guideline\cite{kovach2011blur} was developed by a former Washington bureau chief of The New York Times and a former executive director of the American Press Institute. Drawing on newsroom practices and journalistic craft, they created a practical resource to help the public assess truthfulness through critical questioning amid today's information overload.}
Three key criteria included in this guideline were \edit{\textbf{(1)} source credibility -- considering who or what the sources are and why they are believable; \textbf{(2)} availability of evidence -- assessing what evidence is presented and if it was tested or vetted; \textbf{(3)} consideration of alternative explanations -- reflecting on what an alternative explanation or understanding might exist.} 
We incorporated these criteria by instructing the models to write a detailed fact-checking article for the given claim while considering the credibility of sources for the claim, any evidence that supports the information, and any alternative explanation~\footnote{\edit{These guidelines reflect common practices in health communication, where credible sources often refer to domain experts or scientific studies, validated evidence usually involves empirical or clinical data, and consideration of alternative explanations includes acknowledging both the benefits and drawbacks of certain treatments or policies. In the prompts, we stuck to the general guidelines instead of providing specific examples to avoid biasing LLM outputs.}}. 
Our prompt contained two subtasks (Table~\ref{tab:factchecking_prompts}): classify the veracity of a given claim and provide reasoning for the classification.

Based on this prompting strategy, we collected responses from the LLMs across five runs, ensuring random selection, ordering, and alteration of the few-shot examples given to the models. Following common settings in few-shot learning~\cite{parnami2022learning, mittal2025myth}, we used one true example and one false example for each run from a total of 10 few-shot examples. \edit{While claims from FNIR and FakeNews were all false claims, we included both true and false examples in the prompt to avoid biasing the LLMs toward always treating claims as false in the correctness subtask.}
This setup ensured that each evaluation included balanced guidance for the model in determining claim veracity. 

\edit{After randomly selecting from the outputs of GPT-4, Llama-2-70B, and Mistral-7B to ensure balanced representation across models,} the final dataset comprised 1,498 articles in the zero-shot variation and another 1,498 articles in the few-shot variation (referred to as \Ct{}) with an average length of 335 words (range [9-2135]). \edit{In preliminary prompt testing, we observed differences in article length, which is why our final prompt explicitly requested pieces of `around 800 words.' However, despite this instruction, the LLMs consistently produced more concise articles.}





\subsection*{Communication Styles}
\label{framing}

To compare LLM-generated and human-written fact-checking articles, we analyzed communication styles across three key components -- information, sender, and receiver~\cite{schiavo2013health} -- by examining linguistic features, value alignments, and persuasive strategies.


\vspace{-0.5em}
\begin{itemize}[leftmargin=1em]
\setlength{\itemsep}{0.5pt}
\item \textit{Information Linguistic Features:} For information, we examined linguistic features of certainty, cognitive attributes, and readability~\cite{saha2025ai}. Certainty is the confidence expressed in statements, which is critical in discussing healthcare risk in physician-patient communication~\cite{berry2004risk}. We used the model developed by Pei and Jurgens~\cite{pei-jurgens-2021-measuring} to capture both the level and aspects of certainty in text. Cognitive attributes represent human cognitive processing, such as causal and insightful thinking and is essential to understanding. We used a validated psycholinguistic lexicon, Linguistic Inquiry and Word Count (LIWC)~\cite{pennebaker2001linguistic} to measure cognitive attributes in writing, a tool that has been extensively utilized in empirical studies on misinformation and detection research~\cite{jiang2018linguistic, su2020motivations, zhang2020overview}. Lastly, to measure the accessibility of language, we used Automated Readability Index (ARI)~\cite{smith1967automated} and Flesch-Kincaid grade level~\cite{flesch2007flesch} that indicate the grade level required to understand the text, with a lower score meaning the text is easier to read.

\item \textit{Sender Persuasive Strategies:} On the sender side, we evaluated persuasive strategies employed in the articles in achieving communication goals. We analyzed the persuasive tactics employed to influence opinion or belief~\cite{chen2021persuasion, chen2021weaklysupervisedhierarchicalmodelspredicting} based on a previous work that examined the strategies of credibility, evidence, and impact~\cite{chen2021weaklysupervisedhierarchicalmodelspredicting}. This hierarchical weakly-supervised latent variable model leverages partially labeled data to predict persuasive strategies in text at both the document and sentence levels. 


\item \textit{Receiver Value Alignment:} On the receiver side, we assessed value alignment, which directly impacts the effectiveness of health communication~\cite{kreuter2004role}, by analyzing references to social values and moral foundations. To measure human social values, we utilized Schwartz values~\cite{schwartz, van2023differences} to capture social values of security, power, achievement, hedonism, stimulation, self-direction, universalism, benevolence, conformity, and tradition. Then, to analyze moral considerations in text, we used Mformer~\cite{nguyen2024measuringmoraldimensionssocial}, a fine-tuned large language model that assesses moral foundations, such as authority and fairness, as defined by Moral Foundations Theory~\cite{graham2013moral}. 

\end{itemize}

\vspace{-0.5em}
\edit{Despite the length difference between \Tr{} and \Ct{}, our analyses were designed minimize the length impact. For transformer-based analyses (i.e., certainty, persuasive strategies, Schwartz social values, moral foundations) where token limits apply, we used a paragraph-wise approach: each paragraph was processed individually, and the maximum score across paragraphs was taken as the article-level score to indicate the intensity of a given stylistic feature. This ensured that the entire article contributed to analysis while reducing bias from overall length. For count-based measures, values were normalized by length (specifically, LIWC cognitive attributes normalized by total words, readability normalized at both characters per word and words per sentence), which mitigates potential confounds from content length.}

\subsection*{Human Evaluation}

\edit{Due to the length differences observed between \Tr{} and \Ct{}, we first constructed a pool of human-written and LLM-generated article pairs where length would not perceptibly differ in ways that might affect reader preferences. Specifically, we targeted human-written articles within the 250–400 word range, based on the length distribution of LLM-generated articles (mean = 335), resulting in a pool of 167 article pairs without noticeable length differences. The subsets used in the human evaluation had closely matched average and median lengths between human-written and LLM-generated articles towards the same misinformation claims (human subset: mean = 351.1, median = 351.0; LLM subset: mean = 333.5, median = 328.0).}

To examine human preference for communication styles in fact-checking, we recruited 100 participants from the crowd-sourcing research platform Prolific~\footnote{https://www.prolific.com/} and 99 participants completed the survey. \edit{Attention checks were implemented with time tracked for each pair comparison. We accepted data only from participants whose total reading time exceeded 5 minutes. Completed participants were compensated with an hourly rate of \$12 and} Table~\ref{tab:participantdemo} describes participant demographics. We defined the inclusion criteria as adults who can read English and live in America. 
Participants were informed that the study aimed to examine different strategies for writing fact-checking responses to potential misinformation, without knowing the involvement of AI. 
\edit{At the end of the study, participants were informed that half of the articles had been generated by LLMs. Before launching the study on Prolific, we conducted pilot testing within our institute community to evaluate the website and question clarity, which was not included in the final dataset.}
The survey was hosted on a website created for this study and consisted of three parts \edit{described below (the survey instrument can be found in the Supplement)}:

\vspace{-0.5em}
\begin{itemize}[leftmargin=1em]
\setlength{\itemsep}{0.5pt}
\item \textit{Demographics}: We asked participants about age, gender, race/ethnicity, and level of education. 

\item \textit{Article Rating}: Each participant rated randomly selected five pairs of fact-checking articles, with the order of LLM and human content shuffled. \edit{Each pair consisting of one human-written and one LLM-generated article addressing the same misinformation claim.} For each article pair, without informing the involvement of AI generations and with the order of human-AI article pairs randomized, participants answered three rating questions on a seven-point scale. The three rating questions were: Which article uses more clear and understandable language? Which article provides all the necessary information to understand the fact being checked? Which article do you think is more persuasive?

\item \textit{Preference Reason}: Additionally, we included one open-ended question for each pair of articles about why they preferred a certain article (or the lack of preference). Participants were encouraged to provide at least two factors \edit{explaining their preference for one article over another, or the lack thereof}.
\end{itemize} 					

\vspace{-0.5em}
To examine the effects of demographic characteristics and relative communication style differences on preferences, we performed \edit{cumulative link mixed models (CLMMs) with random intercepts for participant and article-pair}, with outcome variables of preferences in language clarity, information completeness, and persuasiveness. To help explain our regression findings, we also analyzed participants' reasons for their preferences toward LLM-generated articles. Following the general inductive approach~\cite{thomas2006general}, two authors first independently engaged in a close reading of all participant responses to gain a preliminary understanding while taking low-level notes to capture factors mentioned in answers using their own words (e.g., \textit{``facts in everyday terms''}) or quoting participant responses (e.g., \textit{``didn't overwhelm you with information not pertinent to the story''}). Then, they discussed their notes and grouped low-level notes into higher-level themes, such as \textit{``neutrality in presenting information''}. Throughout the analysis, the research team engaged in ongoing discussions to refine and clarify emerging themes.


\section*{Results}
\subsection*{Correctness}

We conducted two sets of prompt experiments: a zero-shot prompt to evaluate the models' inherent ability and style, and a few-shot prompt where we provided human examples of fact-checking articles to help LLMs capture phrasing strategies used by humans. 
The few-shot prompting performed slightly better with 93.68\% in recall, gaining a 2.95\% improvement over zero-shot. 


\subsection*{Communication Styles}

\begin{figure*}[]
\centering
    \begin{subfigure}[b]{0.3\textwidth}
    \centering
    \includegraphics[width=\textwidth]{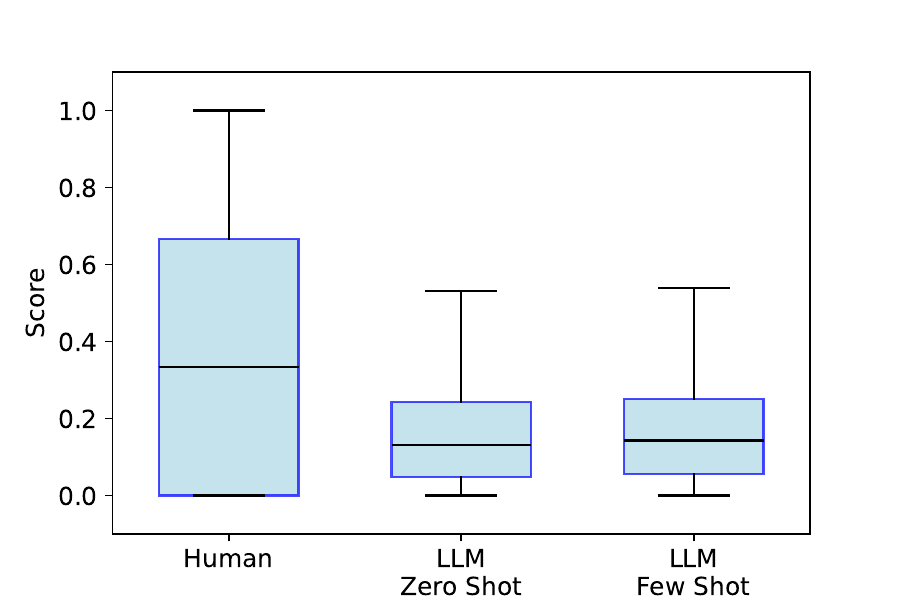}
    \caption{Persuasion: Impact} \label{fig:Impact_persuasion}
    \end{subfigure}
    \hfill
    \begin{subfigure}[b]{0.3\textwidth}
    \centering
    \includegraphics[width=\textwidth]{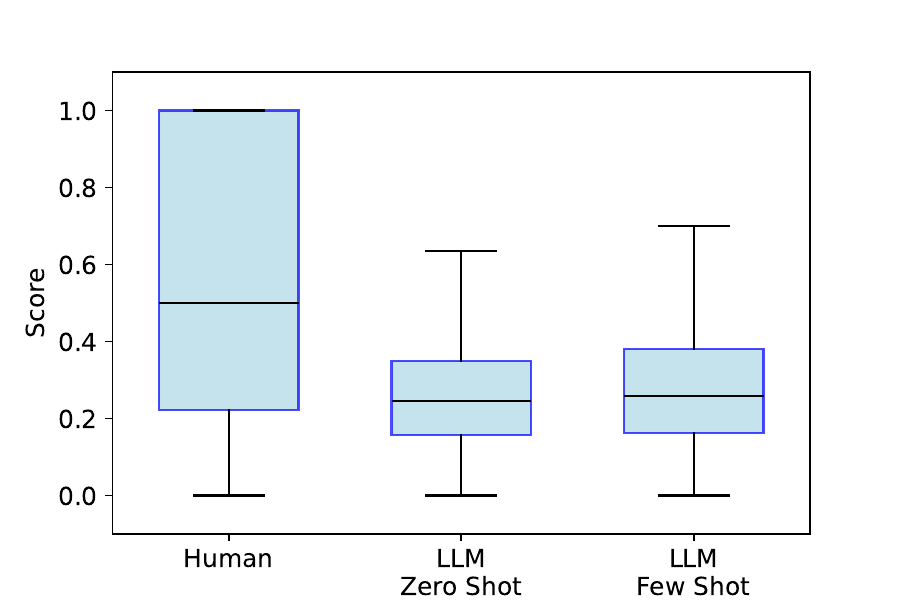}
    \caption{Persuasion: Evidence} \label{fig:Evidence_persuasion}
    \end{subfigure}
    \hfill
    \begin{subfigure}[b]{0.3\textwidth}
    \centering
    \includegraphics[width=\textwidth]{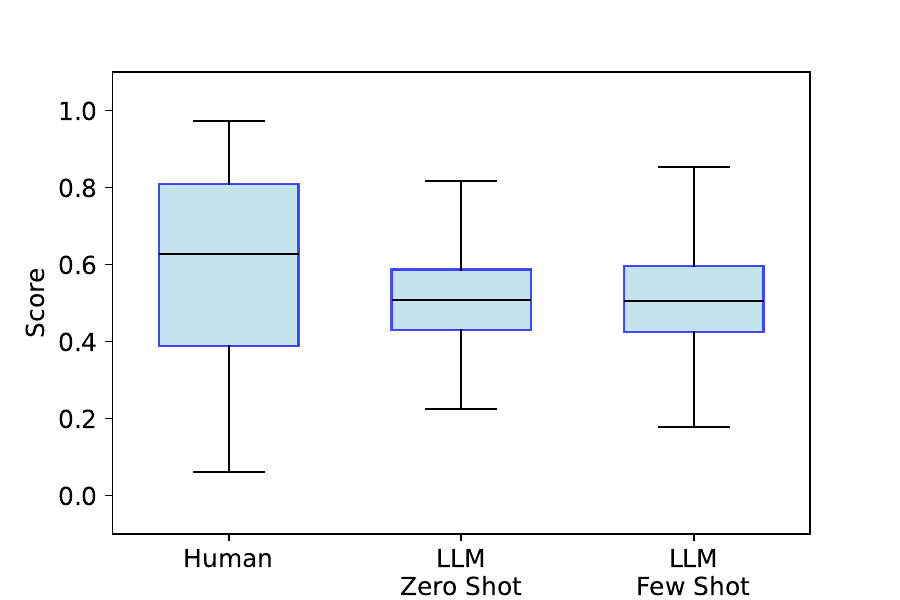}
    \caption{Moral foundation: Fairness} \label{fig:fairness_mf}
    \end{subfigure}

    \begin{subfigure}[b]{0.3\textwidth}
    \centering
    \includegraphics[width=\textwidth]{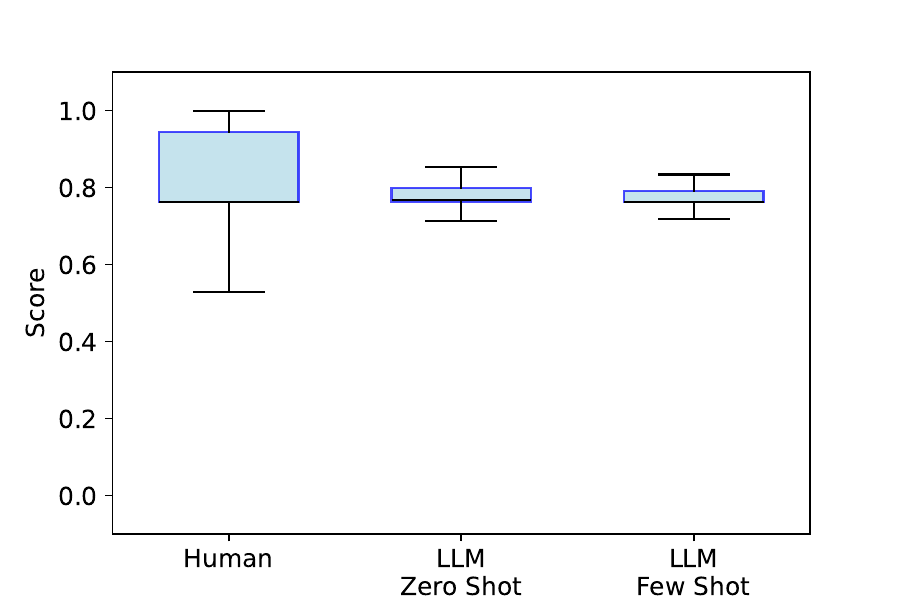}
    \caption{Social value: Power} \label{fig:power_schwartz}
    \end{subfigure}
    \hfill
    \begin{subfigure}[b]{0.3\textwidth}
    \centering
    \includegraphics[width=\textwidth]{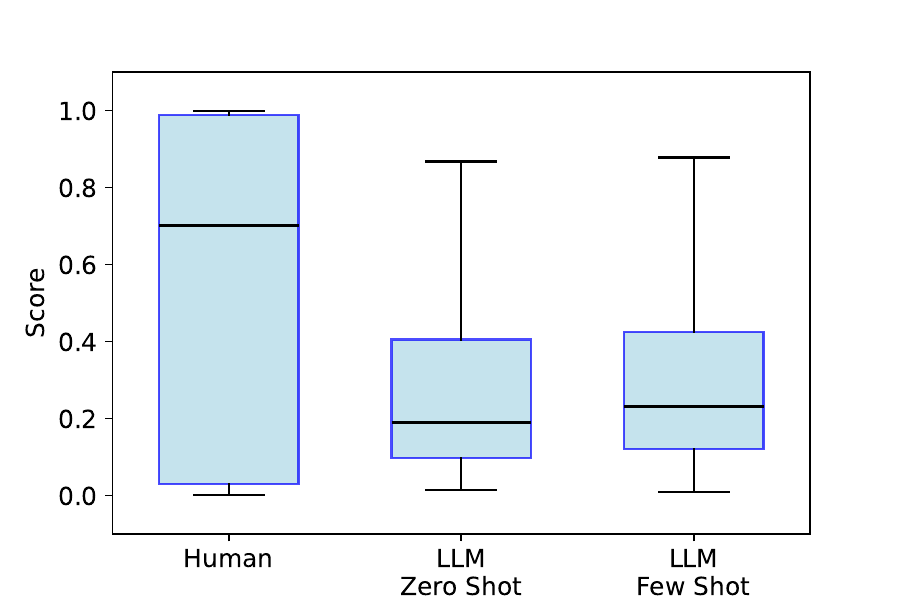}
    \caption{Social value: Conformity} \label{fig:conformity_schwartz}
    \end{subfigure}
    \hfill
    \begin{subfigure}[b]{0.3\textwidth}
    \centering
    \includegraphics[width=\textwidth]{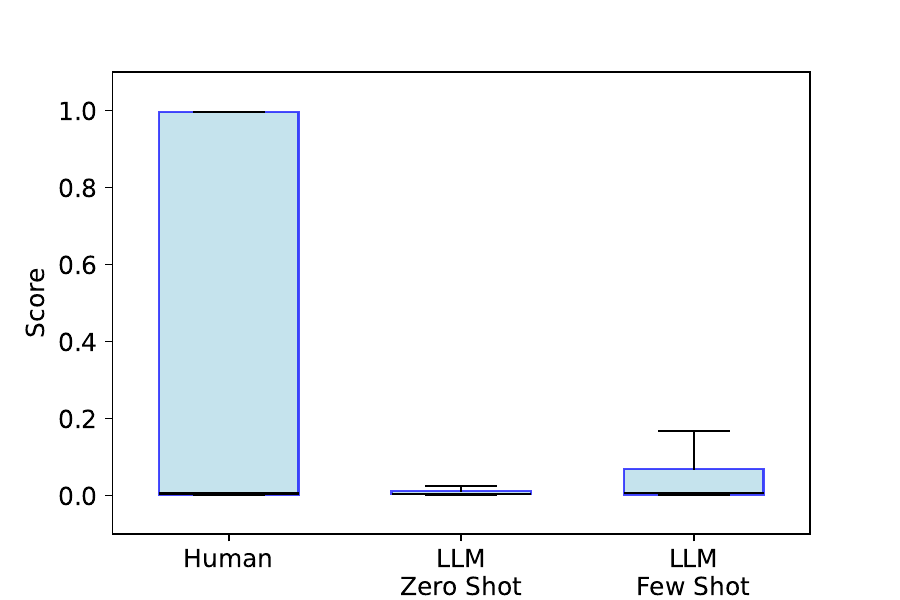}
    \caption{Social value: Tradition} \label{fig:tradition_schwartz}
    \end{subfigure}
    
    \begin{subfigure}[b]{0.3\textwidth}
    \centering
    \includegraphics[width=\textwidth]{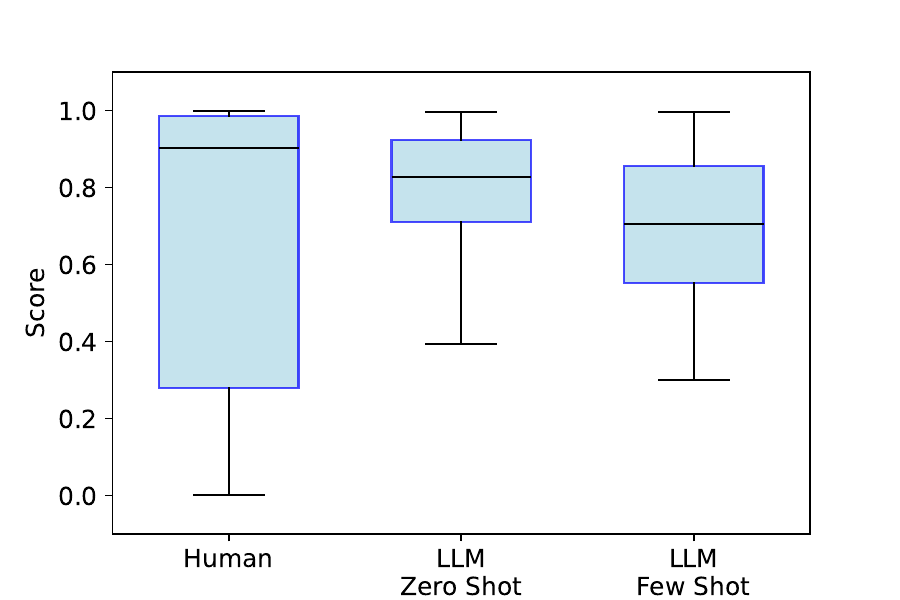}
    \caption{Social value: Achievement} \label{fig:achievement_schwartz}
    \end{subfigure}
    \hfill
    \begin{subfigure}[b]{0.3\textwidth}
    \centering
    \includegraphics[width=\textwidth]{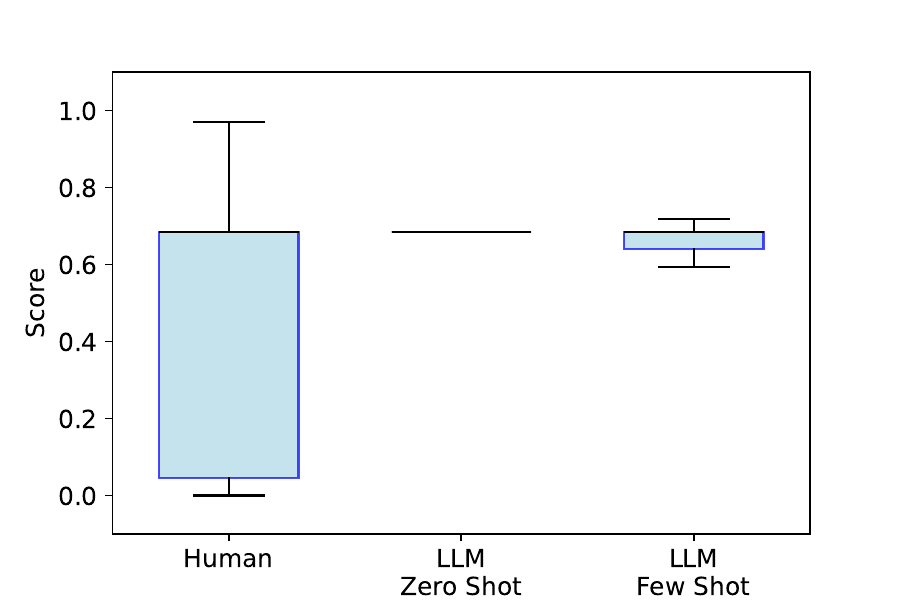}
    \caption{Social value: Stimulation} \label{fig:stimulation_schwartz}
    \end{subfigure}
    \begin{subfigure}[b]{0.3\textwidth}
    \centering
    \includegraphics[width=\textwidth]{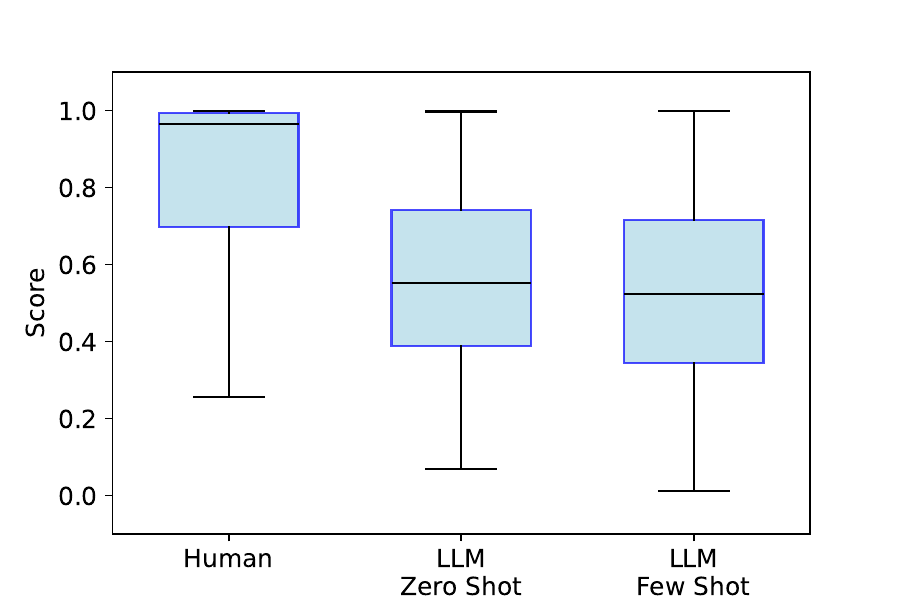}
    \caption{Social value: Self-direction} \label{fig:self_direction_schwartz}
    \end{subfigure}
    
    \caption{Framing dimensions with significant differences between LLMs and human writing}
    \label{fig:linguistics}
\end{figure*}

ANOVA tests showed significant differences for all the communication style dimensions of linguistic features, persuasive strategies, and value alignments (Table~\ref{tab:anova}).
To ensure a robust comparison, we also applied Tukey's HSD test to further analyze these differences and identify specific pairs with statistically significant variations. 
Figure~\ref{fig:linguistics} presents the score distribution plots for both human and LLM assessments. For brevity, LLM scores were averaged across the three LLMs used for this study: GPT-4, Llama-2-70B, and Mistral-7B. Individual models' scores were presented in  Table~\ref{tab:tukeyhsd2a} in Supplementary Materials.

\vspace{0.5em}
\noindent\textbf{\textit{Information linguistic features}}

\noindent\textit{Certainty:}
LLM-generated content scored significantly lower in terms of certainty than human writing, although the mean difference between the groups was smaller as compared to the other dimensions tested.

\noindent\textit{Cognitive attributes:}
LLM-generated content involved significantly more cognitive processing expressions. With higher explanations of cognitive justifications, LLMs could achieve better reasoning in writing. 

\noindent\textit{Readability:}
We found LLM articles had significantly higher readability scores, meaning that they were more complex and harder to read than human writings.
The lower competence of LLM in readability is especially problematic for health claims and misinformation corrections because it may hinder understanding across a diverse audience and weaken trust by failing to present clear and accessible information.

\vspace{0.5em}
\noindent\textbf{\textit{Sender persuasive strategies}}

\noindent LLM-generated articles presented lower language impact and less evidence compared to human-authored articles (Figure~\ref{fig:Impact_persuasion} and \ref{fig:Evidence_persuasion}).
The Tukey HSD test confirmed these observations with a negative mean difference for all LLM-human pairs.

\vspace{0.5em}
\noindent\textbf{\textit{Receiver value alignment}}

\noindent\textit{Social values:}
LLM articles presented lower levels of power (Figure~\ref{fig:power_schwartz}) and achievements (Figure~\ref{fig:achievement_schwartz}) than human writings. 
LLMs' less frequent presentations of control or dominance suggest that human authors tended to be better at establishing authority and guiding readers toward correct information in effective health communication.
LLM's lower emphasis on competence indicates that human authors tended to be skilled at presenting information in a way that aligns with societal expectations of accuracy and reliability. 
We also found that LLM-generated articles exhibited lower tradition (Figure~\ref{fig:tradition_schwartz}) and conformity (Figure~\ref{fig:conformity_schwartz}) scores than human writing. A lower tradition score suggests that LLM content was less culturally grounded, which can affect how people view customs and religions. However, this neutrality can be a strength, helping to avoid cultural bias and promote inclusive content. A lower conformity score suggests that LLMs are less likely to spread biases or stick to cultural stereotypes like human writers might. Lower conformity could be a positive sign, meaning the content is more neutral and credible.

\noindent\textit{Moral foundations:}
LLM content scored significantly lower in moral dimensions than human content, suggesting that LLMs are less likely to emphasize fairness and authority in writing.         

\subsection*{Human Evaluation}

\noindent\textbf{\textit{Reader preference}}

\begin{figure*}[]
\centering
    \includegraphics[width=\textwidth]{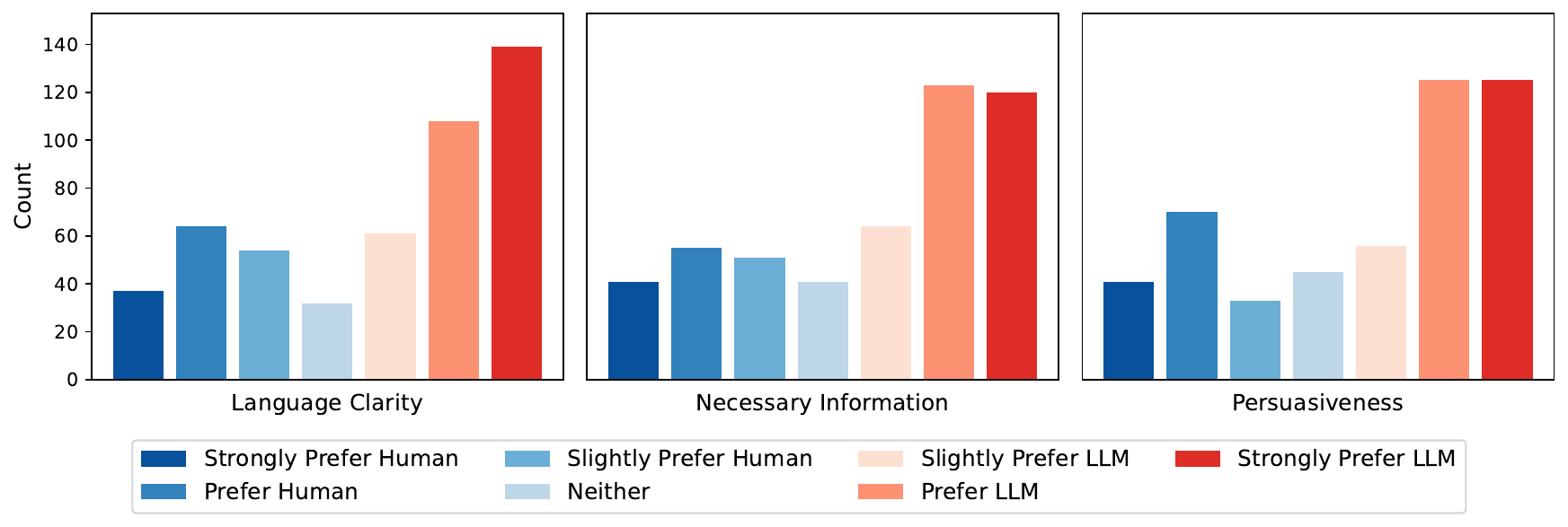}
    \caption{Human preferences for language clarity, necessary information and persuasiveness \edit{between human-written articles and LLM-generated articles. Here LLM articles were generated with few-shot prompts and randomly selected from three models (GPT-4, Llama-2-70B, and Mistral-7B).}}
    \label{fig:human_eval_results}
\end{figure*}

\noindent To assess preference between LLMs and human-generated articles, 99 participants each rated five random content pairs and together provided a total of 495 responses. In general, more individuals preferred LLM-generated articles regarding clarity of language, inclusion of essential information, and persuasiveness. The Wilcoxon Signed-Rank Test rejected the null hypothesis of no preference and showed a significant preference for LLM-generated articles.
Figure~\ref{fig:human_eval_results} shows the human preferences, and regression analyses of individual and content-related factors influencing these preferences are provided in Table~\ref{table:regression}.

\vspace{-0.6em}
\begin{itemize}[noitemsep]
\item \textit{Language clarity:}
 308 out of 495 responses (62.23\%) preferred LLM-generated articles for their clarity and ease of reading, with 28.08\% strongly preferred LLMs. This suggests that LLM content might be better at helping people understand complex health information. Regression results showed that male individuals and individuals identifying as Black/African American and White were more likely to prefer LLM-generated articles. At the same time, people were more likely to prefer articles that expressed more security or tradition-related values or utilized less language impact. 
 We found that LLM articles with lower readability scores (i.e., easier to read) were \edit{significantly more likely to be} rated as clearer, and, while education levels were not a significant factor, participants with higher education levels were more likely to find LLMs clearer.
 
 \item \textit{Necessary information:}
307 out of 495 responses (62\%) preferred LLM-generated articles in providing comprehensive information on health claims. Conversely, a substantial proportion of participants (29.9\%) expressed a preference for human-generated content. 
Regression results indicated that male participants tended to prefer LLM fact-checking and its reasoning style as more thorough in including necessary information.

\item \textit{Persuasiveness:}
306 out of 495 responses (61.8\%) found LLM-generated articles to be more persuasive, with 125 responses (25.25\%) strongly preferring LLMs. 29.09\% participants were inclined towards human-generated articles, with 41 responses (8.28\%) strongly preferring human articles. 
Contrary to prior literature showing authorities as trusted sources with high persuasiveness due to source credibility~\cite{pornpitakpan2004persuasiveness, dutta2003trusted}, our regression results indicated that when LLM articles used fewer authority-related expressions than their paired human articles, participants were more likely to find LLMs more persuasive.
    
\end{itemize}

\vspace{0.5em}
\noindent\textbf{\textit{Reasons for witnessed LLM preference}}

\noindent To understand factors contributing to the observed preference for LLM fact-checked articles, we analyzed participants' reasons for preferences without knowing the involvement of AI generations and summarized five factors:
\vspace{-0.5em}
\begin{itemize}[noitemsep]
    \item \textit{Focused presentation}: Participants preferred LLM-generated fact-checking articles for their clear organization, direct presentation, and focused content -- often starting with the main point and avoiding unnecessary details. The paired human-written ones, on the other hand, were sometimes described as scattered, convoluted, and prone to including excessive or off-topic information, and sometimes giving undue emphasis to false claims.
    \item \textit{Objective standpoint}: LLM-generated fact-checking articles were often characterized as more objective and fact-based with minimal opinions, while human-written ones were sometimes associated as opinionated or personal.
    \item \textit{Professional vibe}: Many participants perceived LLM-generated fact-checking articles as more professional-sounding with cross-checking among multiple organizations, well-explained background information, and references from reputable sources. Comparatively, some human-written articles used \textit{``non-reputable citations''} or \textit{``social media evidence''}.
    \item \textit{Language accessibility}: Participants found LLM-generated fact-checking articles easier to follow with clear and digestible language that involved \textit{``everyday terms to present facts''} in an accessible manner.
    \item \textit{Non-emotional tone}: Participants also pointed out that LLM-generated fact-checking articles employed a calm and neutral tone, while some human-written articles were perceived as \textit{``angry''} or \textit{``accusatory''} and sometimes resembling clickbait grabbing attention.
\end{itemize}

\section*{Discussion}

\edit{While factual correctness is the foundation of health communication, accuracy alone does not guarantee that information will be understood or accepted; instead, communication styles influence the effective delivery of accurate information. Therefore, our study examines the potential of LLMs as LLMs as assistive tools for professionals in clarifying health information} and presents empirical evidence on the communicative features and human preferences between LLM-generated and human-written fact-checking articles addressing health misinformation. Language analysis indicated some potential limitations in LLM-generated content compared to human-written ones. Specifically, LLM-generated fact-checking articles \edit{showed lower scores on indicators of} persuasive strategies in language impact, credibility, and evidence; displayed less certainty \edit{expressions} in writing; and \edit{exhibited reduced alignment on linguistic proxies for} moral foundations of authority and fairness as well as important social values like tradition, self-direction, and conformity. These findings suggest that LLMs \edit{communication styles} may prioritize writing-level persuasion over reasoning support and moral alignment. In addition, LLM articles tended to be more challenging to read for people without high education levels. Although they \edit{presented more linguistic indicators associated with} cognitive processing explanations, this characteristic could potentially make the text feel too analytical rather than intuitive and accessible.

Despite these potential limitations of LLMs presented by language analysis, our human evaluation results suggested a stronger preference towards LLM articles in \edit{perceived} writing clarity, inclusion of necessary information, and persuasiveness. Although this preference for clarity contrasted with the language analysis, our qualitative findings suggested that readers appreciate LLMs' use of digestible terms and their structured and focused presentation style, even if readability metrics indicate otherwise. Similarly, we were surprised to find that human preferences in informational completeness and persuasiveness also contradicted the language analysis. Specifically, LLMs \edit{showed lower scores on linguistic indicators of} evidence use and \edit{expressions associated with authority and fairnessm, while exhibited reduced alignment on proxies for social values} --- dimensions that typically contributed to the comprehensiveness of information and effectiveness of persuasion~\cite{thon2017believing}. However, qualitative results suggested that LLMs' tendency to cite reputable sources, even when citations are incorrect or non-existent, contributed to a perception of professionalism and evidence support.
Together, our findings suggest that people may prioritize surface-level information presentation --- such as structured and focused answers, an objective and neutral tone, and a professional appearance --- over characteristics traditionally associated with high-quality fact-checking and health communication like nuanced reasoning, moral alignment, and holistic coverage.

\vspace{0.05in}\noindent\textbf{LLMs for Fact-checking Explanation: } 
Taken together, our findings suggest the potential of leveraging LLMs in fact-checking workflows while balancing their strengths and limitations. Human preference towards LLM articles despite opposite findings from the language analyses implies that the way LLMs present information may create an impression of completeness and persuasiveness, even if the content itself may fall short on depth or rigor. Thus, while LLMs should not replace human fact-checkers due to their inferior performance on credibility and evidence and weaker alignment with moral and social values, we see potential in them complementing human efforts by presenting and organizing information in a recipient-friendly manner. Future work can study how to fine-tune models to enhance factors valued by high-quality fact-checking such as evidence support and value alignment. 

\vspace{0.05in}\noindent\textbf{LLMs in \edit{Professional-led} Health Communication: } 
In a broader sense, our work sheds light on the potential role of LLMs in assisting health communication. The observed human preference for LLM-generated content in language clarity, information completeness, and persuasiveness suggests the potential for using LLMs in organizing and revising patient-facing materials. However, their weaker performance on reasoning support and moral alignment could pose risks in health contexts where cultural sensitivity, trust-building, and reliability are critical~\cite{schiavo2013health}. These limitations underscore the need for careful evaluation and responsible use of LLMs in high-stakes contexts like healthcare and combating misinformation. Future research can explore developing strategies to fine-tune and monitor LLMs while implementing human oversight to ensure reliability, accountability, and ethical alignment in their applications.

\subsection*{Limitations}

From a generalization standpoint, this study is situated in the context of fact-checking explanations for health information within the U.S. environment. While our findings offer insights into communication style differences and reader perceptions in LLM-generated content, they may not directly translate to other types of health communication tasks. Nonetheless, we believe the broader insight that LLMs lack alignment with professional standards yet appeal to readers through structured and neutral presentation holds relevance across domains.
Second, we choose to study fact-checking articles, as they reflect key elements in health communication: evidence support, trust-building, and support for audience sensemaking. Our evaluation assumes that authoritative organizations' fact-checking articles represent professional styles. While these serve as valuable reference points, they are not definitive gold standards in all health communication contexts. 
\edit{On one hand, while fact-checkers are trained in evidence-based reporting and heavily rely on clinical and scientific evidence, they are not biomedical experts. On the other hand, fact-checking} quality and style can vary by goals, organization, and topic, and may introduce variability in our alignment assessments. As such, our analysis is better interpreted as a relative comparison rather than an absolute measure of style or quality. \edit{Similarly, we do not claim our approach to be a comprehensive or exhaustive examination of all potential dimensions of health communication. For example, we did not evaluate qualities such as coherence or informativeness. The communication style features examined in this study were theory-guided but should be interpreted as proxies rather than definitive measures of communication style. For human study, while randomization helped distribute model types across participants at the cohort level, we cannot ensure that model exposure was balanced within participants. Lastly, this study did not evaluate the factual quality of LLM-generated outputs. Our analyses focused solely on communication styles and related reader preferences to examine the potential of LLMs as assistive writing tools for health communication. As a result, while we acknowledge factual accuracy or completeness as the foundation of health communication, participants' preferences in this study should only be interpreted at the level of communicative appeal rather than factual reliability.}



\section*{Conclusion}
Our study highlights a disconnection between expert-derived communication benchmarks and lay reader preferences in the context of LLM-generated \edit{COVID-19} information explanations. Despite scoring lower on traditional \edit{communication style} dimensions of high-quality health communication, such as persuasive strategies, certainty expression, and alignment with moral and social values, LLM-generated fact-checking responses were consistently preferred by participants. This preference appears to be driven by LLMs’ structured presentation, clarity, and neutral tone, which may foster an impression of completeness and professionalism, even when the underlying content lacks nuance or rigor. Together, these findings indicate both the potential and the limitations of deploying LLMs in fact-checking and health communication workflows. While LLMs can enhance the accessibility of complex information, their current shortcomings in value-sensitive communication raise important concerns for use in high-stakes settings. Future research should focus on fine-tuning LLMs to improve alignment with ethical and communicative standards in health, and on developing frameworks for integrating LLMs responsibly alongside human expertise.

\section*{List of abbreviations}
LLM: large language model; AI: artificial intelligence; CLMM: cumulative link mixed model; LIWC: linguistic inquiry and word count;  ARI: automated readability index; ANOVA: analysis of variance; HSD: honestly significant difference; CI: confidence interval

\clearpage
\section*{Declarations}

\subsection*{Ethics approval and consent to participate}
This study was considered exempt with minimal risk by the Institutional Review Board of Georgia Institute of Technology. No personal information was collected, and participants were informed that the presented statements contained misinformation, followed by fact-checking explanations.

\subsection*{Consent for publication}
Not applicable.

\subsection*{Availability of data and materials}
The datasets are available from the corresponding author on reasonable request.

\subsection*{Competing interests}
The authors declare that they have no competing interests

\subsection*{Funding}
JZ, MC, and MDC were partly supported by U.S. National Science Foundation grant \#2137724. This research project has benefited from the Microsoft Accelerating Foundation Models Research (AFMR) grant program.

\subsection*{Authors' contributions}
JZ and KV contributed to study conceptualization and design, data collection and analysis, result interpretation, and manuscript writing. MC contributed to study conceptualization and design, and data collection and analysis. KS contributed to study conceptualization, result interpretation, and manuscript writing. MDC contributed to and supervised study conceptualization and design, result interpretation, and manuscript writing. All authors read and approved the final manuscript.

\subsection*{Acknowledgements}
We thank the study participants for contributing to this work.

\clearpage 
\section*{Tables and Figures}
\begin{table}[H]
\centering
\footnotesize
\renewcommand{\arraystretch}{1.2}
\begin{tabular}{p{0.95\linewidth}}
\textbf{(a) Claim Verification (Boolean)} \\
\begin{tcolorbox}[colback=gray!15, colframe=gray, boxrule=0.2mm, arc=1mm, left=2pt, right=2pt, top=2pt, bottom=2pt]
\footnotesize
\begin{verbatim}
Let's think step by step.
Establish the credibility of sources and return TRUE or FALSE for the given claim.
[MASK]
<claim-to-verify>
\end{verbatim}
\end{tcolorbox}

\textbf{(b) Fact-Checking Article Generation} \\
\begin{tcolorbox}[colback=gray!15, colframe=gray, boxrule=0.2mm, arc=1mm, left=2pt, right=2pt, top=2pt, bottom=2pt]
\footnotesize
\begin{verbatim}
Write a detailed fact-checking article of around 800 words for the given claim. 
Give a detailed explanation.
Let's think step by step. 
What are the sources and their credibility?
Is there any evidence that supports the information or claim? 
What might be an alternative explanation or understanding?
[MASK]
<claim-to-verify>
<classification-result>
\end{verbatim}
\end{tcolorbox}
\end{tabular}
\caption{Prompt templates for fact-checking tasks, with the [MASK] token serving as a placeholder for few-shot examples.}
\label{tab:factchecking_prompts}
\end{table}

\begin{table*}[h]
\centering
\footnotesize
\begin{tabular}{lr}
\textbf{Demographic} & \textbf{Number of participants} \\
\toprule
\textit{\textbf{Age Distribution}} & \\
18-24 & 10 \\
25-34 & 31 \\
35-44 & 28 \\
45-54 & 14 \\
55-64 & 9 \\
65 or above & 7 \\ \hline
\textit{\textbf{Gender Distribution}} & \\
Male & 36 \\
Female & 62 \\
Prefer not to say & 1 \\ \hline
\textit{\textbf{Ethnicity Distribution}} & \\
American Indian or Alaska Native & 0 \\
Asian & 7 \\
Black or African American & 23 \\
Hispanic or Latino or Spanish Origin & 9 \\
Native Hawaiian or Pacific Islander & 2 \\
White & 53 \\
Other & 5 \\ \hline
\textit{\textbf{Education Distribution}} & \\
High School or less & 13 \\
Some college or technical certification & 28 \\
Bachelor’s degree & 39 \\
Master’s degree or higher & 19 \\
Prefer not to say & 0 \\
\bottomrule
\end{tabular}
\caption{Overview of participant demographics (N = 99).}
\label{tab:participantdemo}
\end{table*}

\begin{table*}[]
\centering
\footnotesize
\begin{tabular}{lll}
\textbf{Dimension} & \textbf{F-statistic} & \textbf{p-adj} \\
\toprule
\textit{\textbf{Persuasive Strategies}} & & \\
Impact & $303.87$ & *** \\ 
Credibility & $133.57$ & *** \\
Evidence & $518.88$ & *** \\ \hline
\textit{\textbf{Certainty}} & & \\
Certainty & $63.97$ & *** \\ \hline
\textit{\textbf{Moral Foundations}} & & \\
Authority & $51.02$ & *** \\
Fairness & $94.32$ & *** \\ \hline
\textit{\textbf{Social Values}} & & \\
Security & $72.34$ & *** \\
Power & $154.38$ & *** \\
Benevolence & $53.85$ & *** \\
Conformity & $216.51$ & *** \\
Tradition & $464.75$ & *** \\
Achievement & $65.93$ & *** \\
Hedonism & $167.75$ & *** \\
Stimulation & $361.72$ & *** \\
Self-direction & $277.42$ & *** \\
Universalism & $146.51$ & *** \\ \hline
\textit{\textbf{Readability}} & & \\
ARI & $806.21$ & *** \\
Flesch-Kincaid & $996.88$ & *** \\ \hline
\textit{\textbf{Cognitive Processes}} & & \\
Cogproc & $618.82$ & *** \\
\bottomrule
\end{tabular}
\caption{ANOVA results for framing of reasoning dimensions (*** $p<0.001$, ** $p<0.01$, * $p<0.05$)}
\label{tab:anova}
\end{table*}


\begin{landscape}
\begin{table}[]
    \centering
    \footnotesize
    \begin{tabular}{l|ccc|ccc|ccc}
    & \multicolumn{3}{c|}{\textbf{Language clarity}} 
    & \multicolumn{3}{c|}{\textbf{Necessary information}} 
    & \multicolumn{3}{c}{\textbf{Persuasiveness}} \\
    \cline{2-10}
    \textbf{Variable} 
     & \textbf{Odds Ratio} & \textbf{$p$} & \textbf{95\% CI} 
     & \textbf{Odds Ratio} & \textbf{$p$} & \textbf{95\% CI} 
     & \textbf{Odds Ratio} & \textbf{$p$} & \textbf{95\% CI} \\
    \hline
    
    \textit{\textbf{Demographic characteristics}} & & & & & & \\
    
    Age  & 1.029 &   & (0.907-1.167) & 1.092 & & (0.964-1.237) & 1.056 &   & (0.930-1.198) \\
    Education  & 1.151 &   & (0.936-1.414) & 1.12 & & (0.914-1.374) & 1.166 &   & (0.948-1.434) \\
    Gender (Masculine) & 1.48 &  *  & (1.048-2.092) & 1.482 & * & (1.051-2.090) & 1.183 &   & (0.837-1.671) \\
    Gender (Prefer not to say) & 1.583 &   & (0.328-7.639) & 3.096 & & (0.470-20.409) & 2.514 &   & (0.483-13.082) \\
    Ethnicity (Black/African American)  & 2.055 &  *  & (1.055-4.001) & 1.716 & & (0.894-3.292) & 1.683 &   & (0.873-3.246) \\
    Ethnicity (Hispanic/Latino/Spanish) & 0.984 &   & (0.435-2.227) & 0.668 & & (0.295-1.517) & 1.022 &   & (0.456-2.290) \\
    Ethnicity (Native Hawaiian/Pacific Islander) & 0.628 &   & (0.199-1.987) & 0.394 & & (0.119-1.301) & 0.443 &   & (0.132-1.487) \\
    Ethnicity (Other) & 1.095 &   & (0.468-2.561) & 1.121 & & (0.480-2.620) & 1.335 &   & (0.567-3.145) \\
    Ethnicity (White) & 2.124 &  *  & (1.132-3.983) & 1.041 & & (0.567-1.913) & 1.265 &   & (0.685-2.337) \\
    
    \textit{\textbf{Relative linguistic differences}} & & & & & & \\                                 
  Cognitive process & 1.024 & & (0.867-1.210) & 0.883 & & (0.748-1.041) & 0.915 & & (0.776-1.078) \\
  Social Value: security  & 1.23  &  *  & (1.024-1.478) & 1.062 & & (0.885-1.275) & 1.129 & & (0.935-1.363) \\
  Social Value: power   & 1.059 & & (0.871-1.287) & 1.082 & & (0.892-1.314) & 1.034 & & (0.849-1.260) \\
  Social Value: achievement & 0.894 & & (0.755-1.059) & 1.078 & & (0.910-1.277) & 0.978 & & (0.825-1.159) \\
  Social Value: hedonism  & 0.99  & & (0.857-1.144) & 1.068 & & (0.915-1.245) & 1.085 & & (0.925-1.272) \\
  Social Value: stimulation & 1.047 & & (0.883-1.240) & 1.024 & & (0.863-1.214) & 1.067 & & (0.901-1.263) \\
  Social Value: universalism  & 0.929 & & (0.757-1.140) & 0.878 & & (0.715-1.079) & 0.966 & & (0.792-1.180) \\
  Social Value: benevolence & 1.017 & & (0.851-1.215) & 1.201 & & (0.977-1.477) & 1.136 & & (0.933-1.383) \\
  Social Value: conformity  & 0.905 & & (0.765-1.071) & 0.886 & & (0.756-1.038) & 0.977 & & (0.827-1.155) \\
  Social Value: tradition   & 1.213 &  *  & (1.021-1.441) & 1.133 & & (0.963-1.333) & 1.167 & & (0.990-1.375) \\
  Moral Foundation: authority   & 0.884 & & (0.746-1.047) & 0.847 & & (0.708-1.014) & 0.821 &  *  & (0.690-0.978) \\
  Moral Foundation: fairness  & 1.034 & & (0.849-1.260) & 0.967 & & (0.796-1.175) & 1.009 & & (0.836-1.218) \\
  Certainty & 0.932 & & (0.795-1.093) & 0.924 & & (0.788-1.084) & 0.967 & & (0.829-1.129) \\
  Persuasive Strategy: Credibility  & 0.944 & & (0.771-1.156) & 0.924 & & (0.755-1.131) & 1.039 & & (0.841-1.284) \\
  Persuasive Strategy: Impact   & 0.821 &  *  & (0.681-0.989) & 0.872 & & (0.723-1.052) & 0.985 & & (0.815-1.190) \\
  Persuasive Strategy: Evidence & 1.002 & & (0.848-1.183) & 1.023 & & (0.870-1.203) & 1.065 & & (0.903-1.257) \\
  Readability: ARI  & 1.414 & & (0.935-2.139) & 1.419 & & (0.942-2.138) & 1.244 & & (0.817-1.895) \\
  Readability: Flesch-Kincaid   & 0.658 & * & (0.440-0.986) & 0.686 & & (0.462-1.019) & 0.754 & & (0.499-1.139) \\
    \hline
    \end{tabular}
    \caption{Regression results for reader preferences in language clarity, necessary information provision, and persuasiveness. (*** $p<0.001$, ** $p<0.01$, * $p<0.05$)}
    \label{table:regression}
\end{table}
\end{landscape}

\clearpage
\bibliography{cas-refs}

\clearpage
\section*{Supplement}
\begin{table*}[h!]
    \centering
    \scriptsize
    \begin{tabular}{llcc}
        \textbf{Dimensions} & \textbf{LLM} & \textbf{meandiff} & \textbf{$p$-adj} \\
        \toprule
        \textit{\textbf{Persuasive Strategies}} & & \\
        \multirow{3}{*}{Impact} & GPT4 & -0.2292 & *** \\
        & Llama-2-70B & -0.1957 & *** \\
        & Mistral-7B & -0.1906 & *** \\
        \hdashline[1pt/1pt]
        \multirow{3}{*}{Credibility} & GPT4 & -0.0586 & *** \\
        & Llama-2-70B & 0.0505 & *** \\
        & Mistral-7B & 0.1106 & *** \\
        \hdashline[1pt/1pt]
        \multirow{3}{*}{Evidence} & GPT4 & -0.3104 & *** \\
        & Llama-2-70B & -0.2678 & *** \\
        & Mistral-7B & -0.2418 & *** \\
        \midrule
        
        \textit{\textbf{Certainty}} & & \\
        \multirow{3}{*}{Certainty} & GPT4 & -0.1723 & *** \\
        & Llama-2-70B & -0.0221 & ** \\
        & Mistral-7B & -0.0354 & ** \\
        \midrule

        \textit{\textbf{Moral Foundation}} & & \\
        \multirow{3}{*}{Authority} & GPT4 & 0.064 & *** \\
        & Llama-2-70B & 0.005 & \\
        & Mistral-7B & 0.0025 & \\
        \hdashline[1pt/1pt]
        \multirow{3}{*}{Fairness} & GPT4 & -0.0661 & *** \\
        & Llama-2-70B & -0.0755 & *** \\
        & Mistral-7B & -0.0997 & *** \\
        \midrule

        \textit{\textbf{Social Values}} & & \\
        \multirow{3}{*}{Security} & GPT4 & 0.1061 & *** \\
        & Llama-2-70B & 0.0311 & *** \\
        & Mistral-7B & 0.0571 & *** \\
        \hdashline[1pt/1pt]
        \multirow{3}{*}{Power} & GPT4 & 0.138 & *** \\
        & Llama-2-70B & 0.1041 & *** \\
        & Mistral-7B & 0.125 & *** \\
        \hdashline[1pt/1pt]
        \multirow{3}{*}{Benevolence} & GPT4 & 0.0708 & *** \\
        & Llama-2-70B & 0.0244 & *** \\
        & Mistral-7B & 0.0433 & *** \\
        \hdashline[1pt/1pt]
        \multirow{3}{*}{Conformity} & GPT4 & -0.1789 & *** \\
        & Llama-2-70B & -0.2459 & *** \\
        & Mistral-7B & -0.273 & *** \\
        \hdashline[1pt/1pt]
        \multirow{3}{*}{Tradition} & GPT4 & -0.2538 & *** \\
        & Llama-2-70B & -0.2728 & *** \\
        & Mistral-7B & -0.2715 & *** \\
        \hdashline[1pt/1pt]
        \multirow{3}{*}{Achievement} & GPT4 & 0.09 & *** \\
        & Llama-2-70B & -0.0385 & *** \\
        & Mistral-7B & 0.0151 &  \\
        \hdashline[1pt/1pt]
        \multirow{3}{*}{Hedonism} & GPT4 & -0.0665 & *** \\
        & Llama-2-70B & -0.0639 & *** \\
        & Mistral-7B & -0.066 & *** \\
        \hdashline[1pt/1pt]
        \multirow{3}{*}{Stimulation} & GPT4 & 0.1641 & *** \\
        & Llama-2-70B & 0.1354 & *** \\
        & Mistral-7B & 0.1624 & *** \\
        \hdashline[1pt/1pt]
        \multirow{3}{*}{Self-direction} & GPT4 & -0.1754 & *** \\
        & Llama-2-70B & -0.2972 & *** \\
        & Mistral-7B & -0.2297 & *** \\
        \hdashline[1pt/1pt]
        \multirow{3}{*}{Universalism} & GPT4 & 0.0288 & *** \\
        & Llama-2-70B & 0.0204 & *** \\
        & Mistral-7B & 0.0259 & *** \\
        \midrule
        
        \textit{\textbf{Readability}} & & \\
        \multirow{3}{*}{ARI} & GPT4 & 2.9773 & *** \\
        & Llama-2-70B & 1.9263 & *** \\
        & Mistral-7B & 2.2695 & *** \\
        \hdashline[1pt/1pt]
        \multirow{3}{*}{Flesch-Kincaid} & GPT4 & -2.8112 & *** \\
        & Llama-2-70B & 1.934 & *** \\
        & Mistral-7B & 1.8838 & *** \\
        \midrule

        \textit{\textbf{Cognitive Processes}} & & \\
        \multirow{3}{*}{Cogproc} & GPT4 & 3.2234 & *** \\
        & Llama-2-70B & 3.8885 & *** \\
        & Mistral-7B & 2.9714 & *** \\
        \bottomrule
    \end{tabular}
    \caption{Tukey’s HSD test results between Human and LLMs for Persuasive Strategies, Certainty, Moral Foundations and Social Values (*** $p<0.001$, ** $p<0.01$, * $p<0.05$)}
    \label{tab:tukeyhsd2a}
\end{table*}

\end{document}